\documentclass[3p,times]{elsarticle}

\usepackage{ecrc}

\volume{00}

\firstpage{1}

\runauth{E.~G.~Ferreiro, F.~Fleuret, J.P.~Lansberg, N.~Matagne, A.~Rakotozafindrabe}

\jid{nupha}

\usepackage{amssymb}

\biboptions{sort&compress}

\usepackage[figuresright]{rotating}

\usepackage{amsmath}
\usepackage{color}
\usepackage{graphicx}
\usepackage{float} 
\usepackage{subfig} 
\usepackage{ifpdf}
\ifpdf
\usepackage[pdftex]{hyperref}
\else
\usepackage[hypertex]{hyperref}
\fi

\hypersetup{
  pdftitle={},%
  pdfauthor={},%
  pdfsubject={},%
  pdfkeywords={},%
  pdfstartview={},%
  bookmarksopen=true, breaklinks=true, debug=true, %
  colorlinks=true, linkcolor=red, citecolor=blue, urlcolor=blue
}

\newcommand{\beq}{\begin{eqnarray}}
\newcommand{\eeq}{\end{eqnarray}}
\newcommand{\be}{\begin{eqnarray*}}
\newcommand{\ee}{\end{eqnarray*}}

\newcommand{\ie}{{\it i.e.}}

\newcommand{\cf}[1]{{Fig.~\ref{#1}}}

\def\lsim{\raise0.3ex\hbox{$<$\kern-0.75em\raise-1.1ex\hbox{$\sim$}}}
\def\gsim{\raise0.3ex\hbox{$>$\kern-0.75em\raise-1.1ex\hbox{$\sim$}}}

\def\CuCu {CuCu}

\def\AuAu {AuAu}
\def\pp   {$pp$}
\def\pA   {$pA$}
\def\dA {$dA$}
\def\AA   {$AA$}
\def\AB   {$AB$}

\def\PbPbm {\mathrm{PbPb}}
\def\pPb  {$p$Pb}
\def\pPbm  {p\mathrm{Pb}}

\def\sqrtsNN {\mbox{$\sqrt{s_{NN}}$}}
\def\Npart   {\mbox{$N_{\rm part}$}}
\def\Ncoll   {\mbox{$N_{\rm coll}$}}

\def\RPbPb    {\mbox{$R_{\rm PbPb}$}}

\def\jpsi   {\mbox{$J/\psi$}}
\def\ccbar {\mbox{$c\bar{c}$}}
\def\pT      {\mbox{$P_{T}$}}
\def\kT     {\mbox{$k_{T}$}}
\def\sigabs {\mbox{$\sigma_{\mathrm{abs}}$}}
\def\beq     {\begin{equation}}
\def\eeq     {\end{equation}}

\long\def\symbolfootnote[#1]#2{\begingroup%
  \def\thefootnote{\fnsymbol{footnote}}\footnote[#1]{#2}\endgroup}

\graphicspath{{plots/}}

\begin{document}

%
\begin{frontmatter}



\dochead{}

\title{Cold Nuclear Matter Effects on \jpsi\ Production with Extrinsic \pT\ \\at $\sqrt{s_{NN}}=5.5\mathrm{~TeV}$ at the LHC}


\author[elena]{E.~G.~Ferreiro}
\author[frederic]{F.~Fleuret}
\author[jp]{J.~P.~Lansberg}
\author[nicolas]{N.~Matagne}
\author[andry]{A.~Rakotozafindrabe}

\address[elena]{Departamento de F{\'\i}sica de Part{\'\i}culas, 
Universidad de Santiago de Compostela, 15782 Santiago de Compostela, Spain}
\address[frederic]{Laboratoire Leprince Ringuet, \'Ecole polytechnique, CNRS-IN2P3,  91128 Palaiseau, France}
\address[jp]{IPNO, Universit{\'e} Paris-Sud 11, CNRS/IN2P3, 91406 Orsay, France}
\address[nicolas]{Universit\'e de Mons, Service de Physique Nucl\'eaire et Subnucl\'eaire, Place du Parc 20, B-7000 Mons, Belgium}
\address[andry]{IRFU/SPhN, CEA Saclay, 91191 Gif-sur-Yvette Cedex, France}

\begin{abstract}
Taking into account the gluon shadowing and the $J/\psi$ nuclear absorption, we evaluate the Cold Nuclear Matter (CNM) effects on $J/\psi$ production in $p$Pb and PbPb collisions at the design LHC energy for PbPb collisions, $\sqrt{s_{NN}}=5.5$ TeV. In view of the good agreement between the yield prediction using the LO CSM and the ALICE data at 7 TeV, we employ the latter model to use a complete description of the kinematics of the underlying $2 \rightarrow 2$ partonic process, namely $g + g \rightarrow J/\psi  + g$. We observe a large $J/\psi$ suppression in $p$Pb and PbPb collisions due to the strong shadowing of the gluon distribution in the lead ion at the corresponding gluon momentum fractions. This suppression from CNM effects has a strong rapidity dependence which may compensate, partially or completely, that  of the predicted charm recombination in PbPb collisions.
\end{abstract}


\end{frontmatter}



\section{Introduction}
\label{sec:intro}
Quark Gluon Plasma (QGP), a deconfined state of QCD matter, is expected to be produced by
Relativistic nucleus-nucleus (\AB) collisions at high enough densities or temperatures. The \jpsi\ meson should be~\cite{Matsui86} sensitive to Hot and Dense Matter (HDM) effects, through processes like the colour Debye screening of the \ccbar~pair. A significant suppression of the \jpsi\ yield was observed at SPS energy by the NA50~experiment~\cite{NA50ref}, and at RHIC by the PHENIX experiment in \CuCu~\cite{Adare:2008sh} and \AuAu~\cite{Adare:2006ns} collisions at $\sqrtsNN=200\mathrm{~GeV}$. In 2010,  data have  been taken at the LHC in PbPb collisions at $\sqrt{s_{NN}} = 2.76\mathrm{~TeV}$. After the long LHC shutdown planned in 2012, LHC will hopefully run at its design energy with PbPb collisions at $\sqrt{s_{NN}}= 5.5 \mathrm{~TeV}$. These results will give the possibility to test the available models  at a new energy scale.  However, the interpretation of the results obtained in \AB\ collisions relies on a good understanding and a proper subtraction of the Cold Nuclear Matter~(CNM) effects which are known  to already impact the \jpsi\ production in proton~(deuteron)-nucleus (\pA\ or \dA) collisions, where the deconfinement cannot be reached.
Two CNM effects are of particular importance~\cite{Vogt:2004dh}: (i)  the shadowing of the initial parton distributions (PDFs)
due to the nuclear environment, and (ii) the breakup of \ccbar~pairs
after multiple scatterings with the remnants of the incident nuclei, referred to as the nuclear
absorption. In our previous works~\cite{Ferreiro:2008wc,Ferreiro:2009qr,Ferreiro:2009ur,Rakotozafindrabe:2010su}, we developed an exhaustive study of these effects at $\sqrtsNN=200\mathrm{~GeV}$ and we confronted our results to the measurements from PHENIX~\cite{Adare:2007gn}. 
It is our purpose here to extend our results to \pPb\ and PbPb collisions at the 
energy $\sqrt{s_{NN}} = 5.5\mathrm{~TeV}$. 

\begin{figure*}[htb!]
\vspace{-0.6cm}
\begin{center}

\subfloat[][$pp$ collisions \cite{Lansberg:2010cn}]{%
\label{fig:plot-dsigvsdy-y_0-231010}
\includegraphics[width=0.37\linewidth]{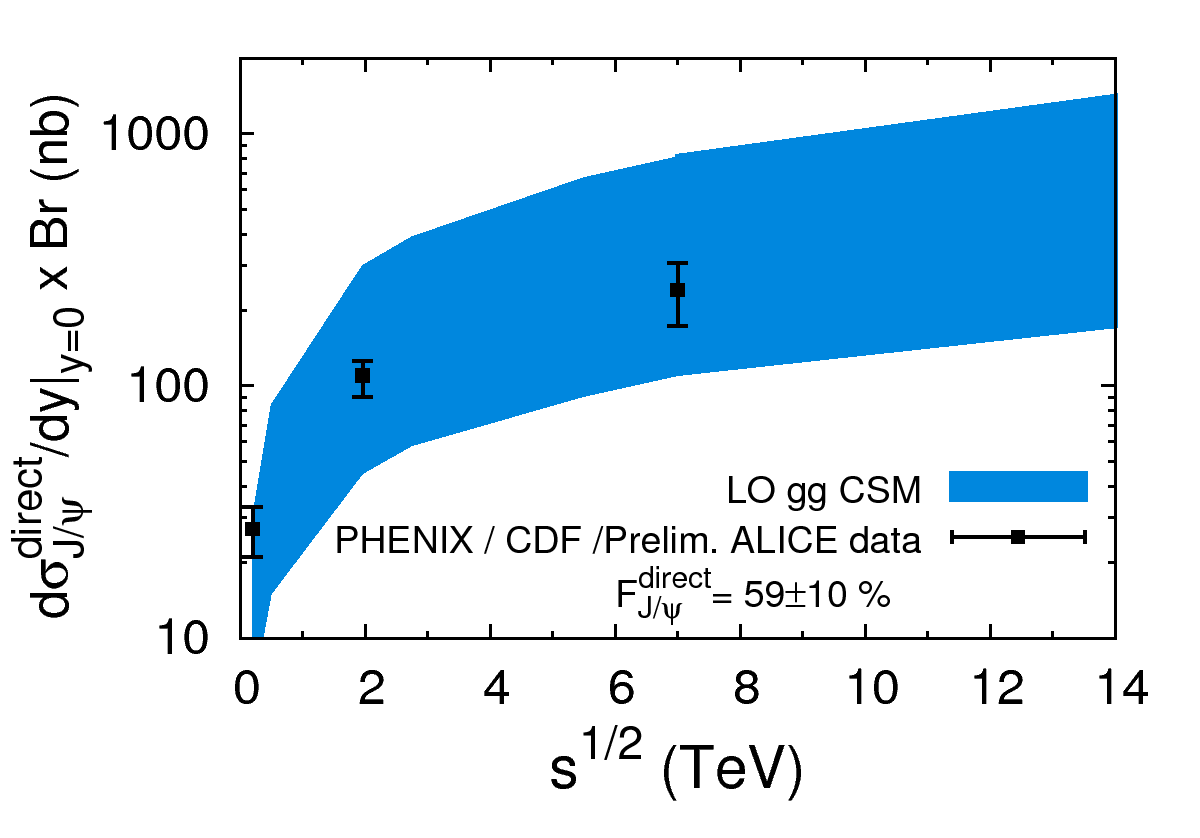}}
\subfloat[][\pPb\ collisions]{%
\label{fig:RpPb_vs_y_and_RPbPb_vs_y_vs_cent-a}
\includegraphics[width=0.34\linewidth]{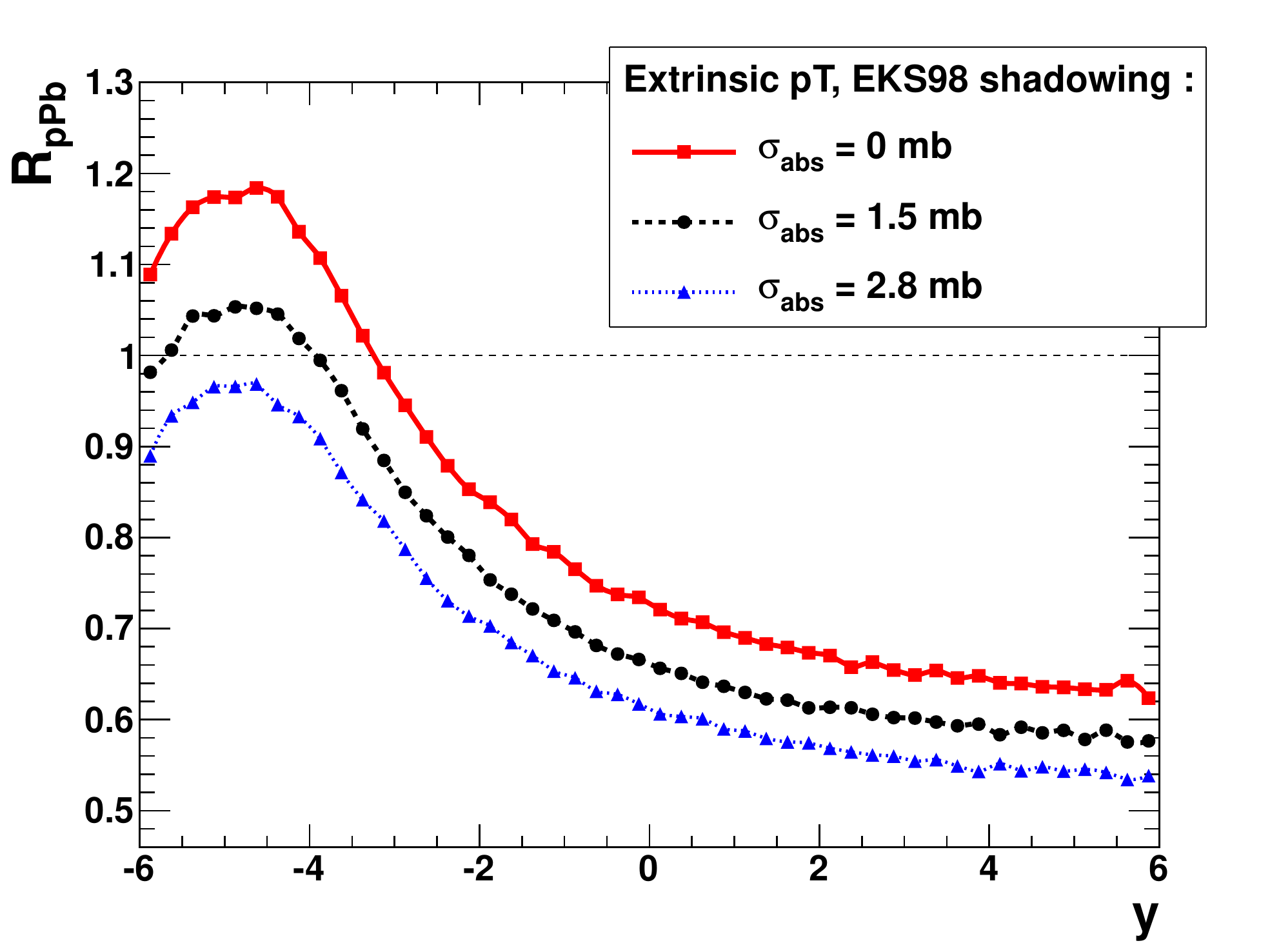}}
\subfloat[][PbPb collisions]{%
\label{fig:RpPb_vs_y_and_RPbPb_vs_y_vs_cent-b}
\includegraphics[width=0.27\linewidth]{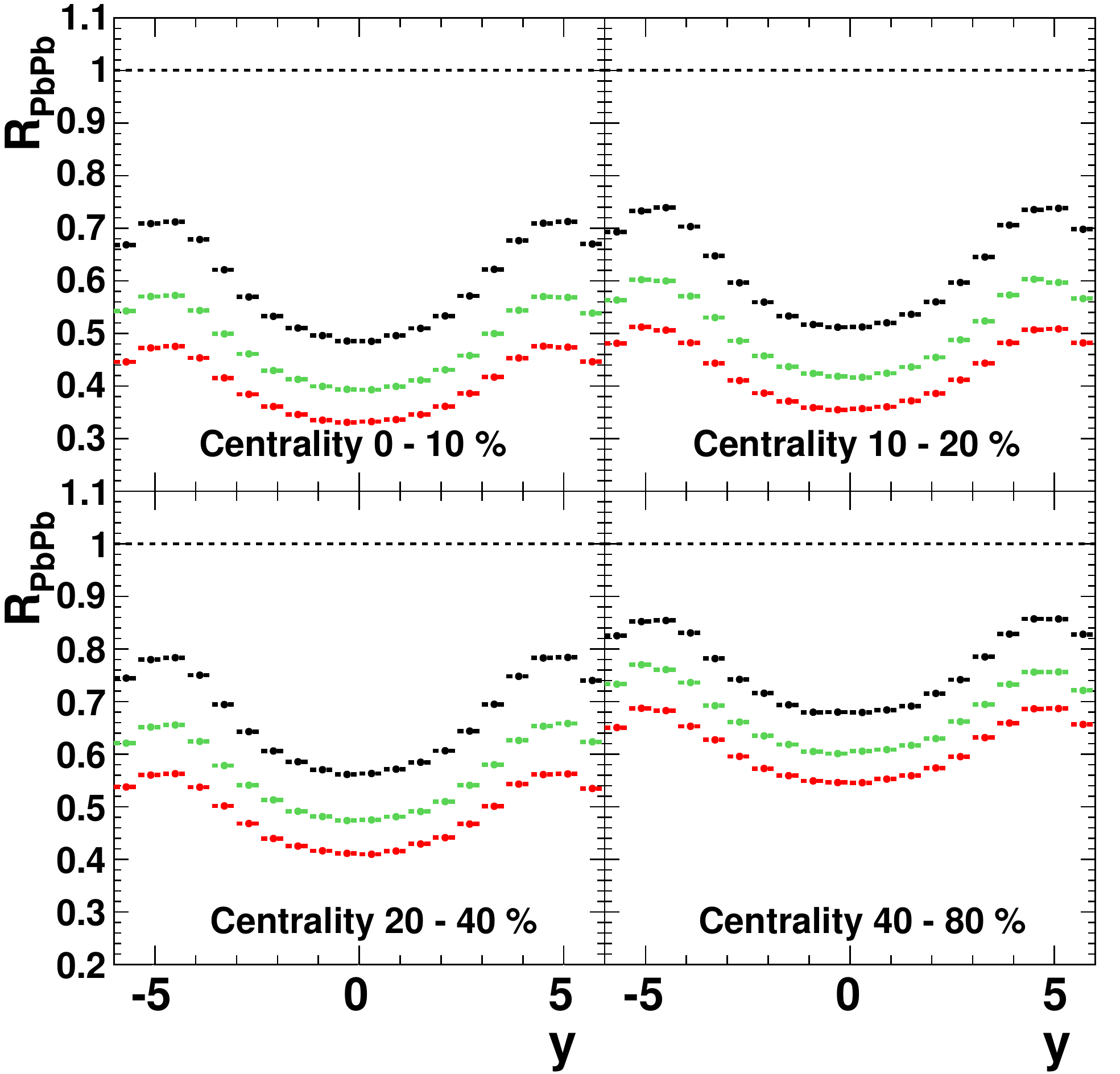}}

\end{center}
\caption{(a) $d\sigma^{\mathrm{direct}}_{J/\psi}/dy|_{y=0}\ \times$ Br from $gg$ fusion in $pp$ collisions for $\sqrt{s}$ from 200 GeV up to 14 TeV compared to the PHENIX \cite{Adare:2006kf}, CDF \cite{Acosta:2004yw} and ALICE data. (b)-(c) \jpsi\ nuclear modification factor versus $y$ in \pPb\ and PbPb collisions at $\sqrt{s_{NN}}=5.5\mathrm{~TeV}$, using EKS98~\cite{Eskola:1998df} gluon shadowing parametrisation and three values of $\sigma_{abs}$ (from top to bottom: $0, 1.5, 2.8\mathrm{~mb}$) in the extrinsic scheme. For PbPb collisions, the $y$-dependence is shown for various centrality selections.}
\label{fig:RpPb_vs_y_and_RPbPb_vs_y_vs_cent}
\end{figure*}

 We have shown in earlier studies~\cite{Ferreiro:2008wc,Ferreiro:2009qr,Ferreiro:2009ur,Rakotozafindrabe:2010su} that \jpsi\ partonic production mechanism  affects both the way to compute the nuclear shadowing and its
expected impact on the \jpsi\ production. Most studies on the \jpsi\ production in hadronic collisions assume that the $c\bar{c}$ pair is produced by a $2 \rightarrow 1$ partonic process where both initial particles are  two gluons carrying
some intrinsic transverse momentum~\kT. The sum of the gluon intrinsic \kT\ is transferred to the \ccbar~pair, thus to
the \jpsi\ since the soft hadronisation process does not 
significantly alter the
kinematics. This is supported by the picture of the Colour Evaporation
Model (CEM) at LO (see~\cite{Lansberg:2006dh} and references 
therein) or of the Colour-Octet (CO) mechanism at
$\alpha_s^2$~\cite{Cho:1995ce}.  Thus, in such approaches, the transverse momentum \pT\ of the
\jpsi\ comes {\it entirely}  from the intrinsic \kT\ of the initial gluons. 
This  is not sufficient to describe the \pT\ spectrum of quarkonia in
hadron collisions \footnote{In addition, recent theoretical
works incorporating QCD corrections or $s$-channel cut contributions have emphasized~\cite{,QCD:recentWorks,Brodsky:2009cf,SchannelCutpapers} that the Colour-Singlet (CS) mediated contributions are sufficient to describe the experimental data for hadroproduction of both charmonium and bottomonium systems without the need of CO contributions. Furthermore, recent works~\cite{ee} focusing on production at $e^+ e^-$ colliders have posed stringent constraints on the size of CO contributions, which are the precise ones supporting a \mbox{$2\to 1$} hadroproduction mechanism~\cite{Lansberg:2006dh}.} \cite{Lansberg:2006dh}. As a consequence, \jpsi\ production at low and mid \pT\ likely proceeds via a \mbox{$2\to 2$} process, such as $g+g \to \jpsi + g$, instead of a \mbox{$2\to 1$} process\footnote{ One may also go further and consider more than two particles in the final state, as expected from the real-emission contributions at NLO and NNLO~\cite{QCD:recentWorks}. It is clear from the yield polarisation~\cite{Lansberg:2010vq} that these contributions start to dominate for $P_T$ above $1-2 m_c$. The effect of more partons in the final state is to increase the difference between the results obtained in both schemes. However the implementation of NLO and NNLO codes in a Glauber model with an inhomogeneous shadowing is not yet available.}. This amounts to the bulk of the \jpsi\ production cross section. For instance, as illustrated by Fig \ref{fig:plot-dsigvsdy-y_0-231010}, the yield predicted by the LO CSM \cite{Lansberg:2010cn} reproduces correctly the PHENIX, CDF and ALICE measurements without resorting to any colour-octet mechanism nor parameter twisting. Consequently, one is entitled to consider that the former \mbox{$2\to 2$} kinematics  is the most appropriate to derive CNM effects at RHIC, and to provide predictions at LHC energy. In this work, we shall focus on the CNM effects expected at the 
energy $\sqrt{s_{NN}} = 5.5\mathrm{~TeV}$ in the extrinsic scheme. The article is organized as follows: in section~\ref{sec:approach},
we will describe our model and in section~\ref{sec:results}, we will present and discuss our results.

\section{Our approach}
\label{sec:approach}

To describe the \jpsi\ production in nuclear collisions, our
Monte~Carlo framework~\cite{Ferreiro:2008wc,OurIntrinsicPaper} is based on the probabilistic Glauber model. The nucleon-nucleon inelastic cross section at
$\sqrt{s_{NN}} = 5.5\mathrm{~TeV}$ is taken to be $\sigma_{NN}=70\mathrm{~mb}$~\cite{Jia:2009mq} and the maximum nucleon density to be $\rho_0=0.17\mathrm{~nucleons/fm}^3$. We also need to implement
the partonic process for the
\ccbar\ production model that allows us to describe the \pp\ data and the CNM effects. 


For $\pT \, \gsim
\ 2-3\mathrm{~GeV}$, most of the transverse momentum of the quarkonia should have an extrinsic
origin, \ie\ the \jpsi's \pT\ would be balanced by the emission of a recoiling particle -- a hard gluon -- in the final
state. 
The \jpsi\ would then be produced by gluon fusion in a \mbox{$2\to 2$} process.
This emission, which is anyhow mandatory to conserve $C$-parity, 
has a definite influence on the kinematics of the
\jpsi\ production. Indeed, for a given \jpsi\ momentum (thus for
fixed rapidity~$y$ and \pT), the processes discussed above, \ie\ the intrinsic $g+g \to \ccbar \to J/\psi \,(+X)$
and the extrinsic $g+g \to J/\psi +g$,  will proceed  from initial gluons with different Bjorken-$x$ on the average. Therefore,
 they will be affected by different shadowing corrections. 

In the intrinsic scheme, the measurement of the \jpsi\ momentum in \pp\ collisions completely fixes the longitudinal
 momentum fraction of the initial partons: $x_{1,2} = \frac{m_T}{\sqrt{s_{NN}}} \exp{(\pm y)} \equiv x_{1,2}^0(y,P_T)$ with $m_T=\sqrt{M^2+P_T^2}$, $M$ being the \jpsi\ mass. On the contrary, in the extrinsic scheme, the knowledge of
the $y$ and \pT\ spectra is not sufficient to determine $x_1$ and $x_2$.
Actually, the presence of a final-state gluon introduces further degrees
of freedom, 
allowing several $(x_1, x_2)$ for a given set $(y, P_T)$.
 The four-momentum conservation 
results in a 
complex expression of $x_2$ as a function of~$(x_1,y,P_T)$:
$
x_2 = \frac{ x_1 m_T \sqrt{s_{NN}} e^{-y}-M^2 }
{ \sqrt{s_{NN}} ( \sqrt{s_{NN}}x_1 - m_T e^{y})} \ .
\label{eq:x2-extrinsic}
$
Equivalently, a similar expression can be written for $x_1$ as a function of~$(x_2,y,P_T)$.
Even if the kinematics determines
the physical phase space, models are anyhow {\it mandatory} to compute the proper
weighting of each kinematically allowed $(x_1, x_2)$. This weight is simply
the differential cross section at the partonic level times the gluon PDFs,
\ie\ $g(x_1,\mu_F) g(x_2, \mu_F) \, d\sigma_{gg\to J/\psi + g} /dy \,
dP_T\, dx_1 dx_2 $.
In the present implementation of our code, we are able to use the partonic differential
cross section computed from {\it any} theoretical approach. In this work, we shall use the Colour-Singlet Model (CSM) at LO at LHC energy, shown to be compatible (see Fig. \ref{fig:plot-dsigvsdy-y_0-231010}) \cite{Brodsky:2009cf,Lansberg:2010cn} with the magnitude of the \pT-integrated cross-section as given by the PHENIX \pp\ data~\cite{Adare:2006kf}, the CDF $p\bar{p}$ data~\cite{Acosta:2004yw} and the LHC \pp\ data at $\sqrt{s_{NN}} = 7\mathrm{~TeV}$. 


To obtain the \jpsi\ yield in \pA\ and \AA\ collisions, a shadowing-correction
factor has to be applied to the \jpsi\ yield obtained from the simple
superposition of the equivalent number of \pp\ collisions.
This shadowing factor can be expressed in terms of the ratios $R_i^A$ of the
nuclear Parton Distribution Functions (nPDF) in a nucleon belonging to a nucleus~$A$ to the
PDF in the free nucleon:
%
$R^A_i (x,Q^2) = \frac{f^A_i (x,Q^2)}{ A f^{nucleon}_i (x,Q^2)}\ , \ \
i = q, \bar{q}, g \ .$
The numerical parameterisation of $R_i^A(x,Q^2)$
is given for all parton flavours. Here, we restrict our study to gluons since, at
high energy, the \jpsi\ is essentially produced through gluon fusion
\cite{Lansberg:2006dh}. Several shadowing parametrisations are available~\cite{Eskola:1998df,Eskola:2009uj}. In the following, we shall restrict ourselves to EKS98~\cite{Eskola:1998df}, which is very close to the mean in the current evaluation of the uncertainty~\cite{Eskola:2009uj} on the gluon nPDF and exhibits a moderate antishadowing. We postpone the propagation of the uncertainty on the gluon nPDF to the CNM effects evaluated at LHC energy for future studies.

The second CNM effect that we are going to take into account concerns
the nuclear absorption.  In the framework of the probabilistic Glauber
model, this effect is usually parametrised
by introducing an effective absorption cross
section~\sigabs. It reflects the break-up of correlated \ccbar~pairs due to inelastic scattering with the remaining nucleons from the incident cold nuclei. The value of~\sigabs\ is unknown at LHC. At high energy, the heavy state in the projectile should undergo a coherent scattering off the nucleons of the target nucleus~\cite{Kopeliovich:2001ee}, in contrast with the incoherent, longitudinally ordered scattering that takes place at low energies. As argued in~\cite{Capella:2006mb,Tywoniuk:2007gy}, this should lead to a decrease of \sigabs\ with increasing $\sqrt{s_{NN}}$. The systematic study of many experimental data indicate that \sigabs\ appears either constant~\cite{Arleo:2006qk} or decreasing~\cite{Lourenco:2008sk} with energy. Hence, we can consider our estimates~\cite{Ferreiro:2009ur,Rakotozafindrabe:2010su} of \sigabs\ at RHIC energy as upper bounds for the value of \sigabs\ at LHC. We choose three values of~\sigabs\ that should span that interval ($\sigma_{\mathrm{abs}} = 0, 1.5, 2.8\mathrm{~mb}$).

\section{Results and discussion}
\label{sec:results}

\begin{figure*}[htb!]
\begin{center}
\subfloat[][$\sigma_{abs}=0\mathrm{~mb}$]{%
\label{fig:RPbPb_vs_Npart-a}
\includegraphics[width=0.3\linewidth]{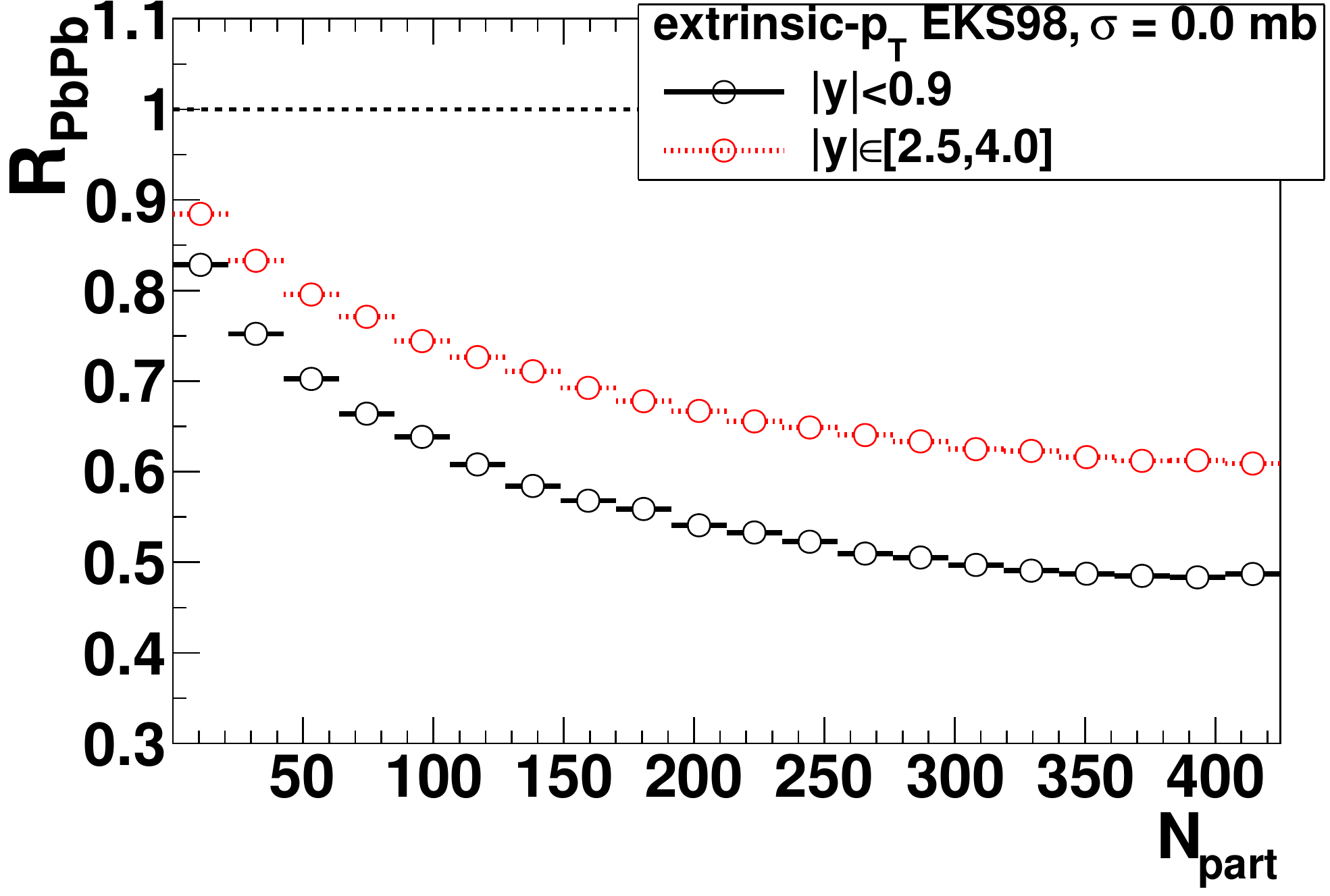}}
\subfloat[][$\sigma_{abs}=1.5\mathrm{~mb}$]{%
\label{fig:RPbPb_vs_Npart-b}
\includegraphics[width=0.3\linewidth]{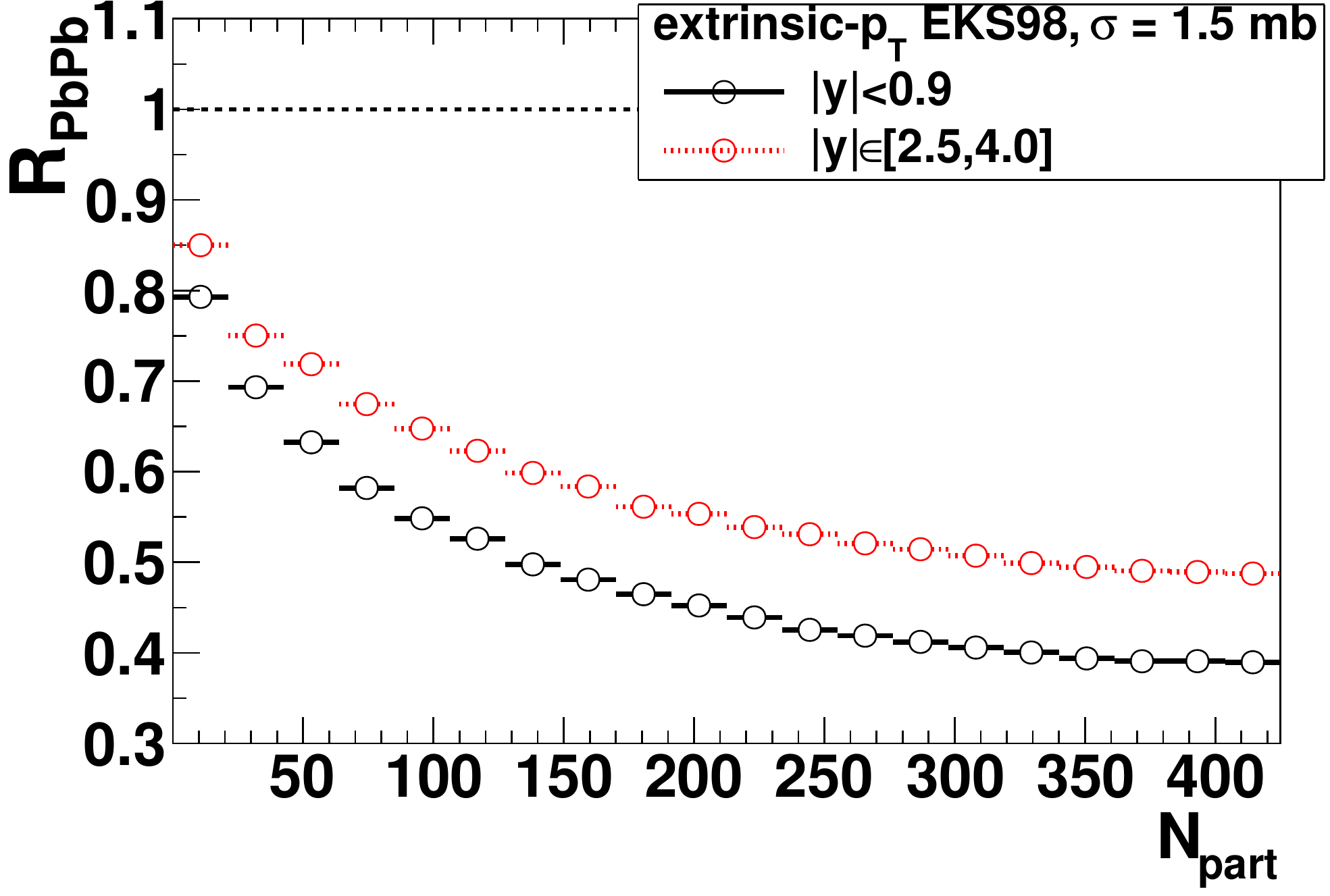}}
\subfloat[][$\sigma_{abs}=2.8\mathrm{~mb}$]{%
\label{fig:RPbPb_vs_Npart-c}
\includegraphics[width=0.3\linewidth]{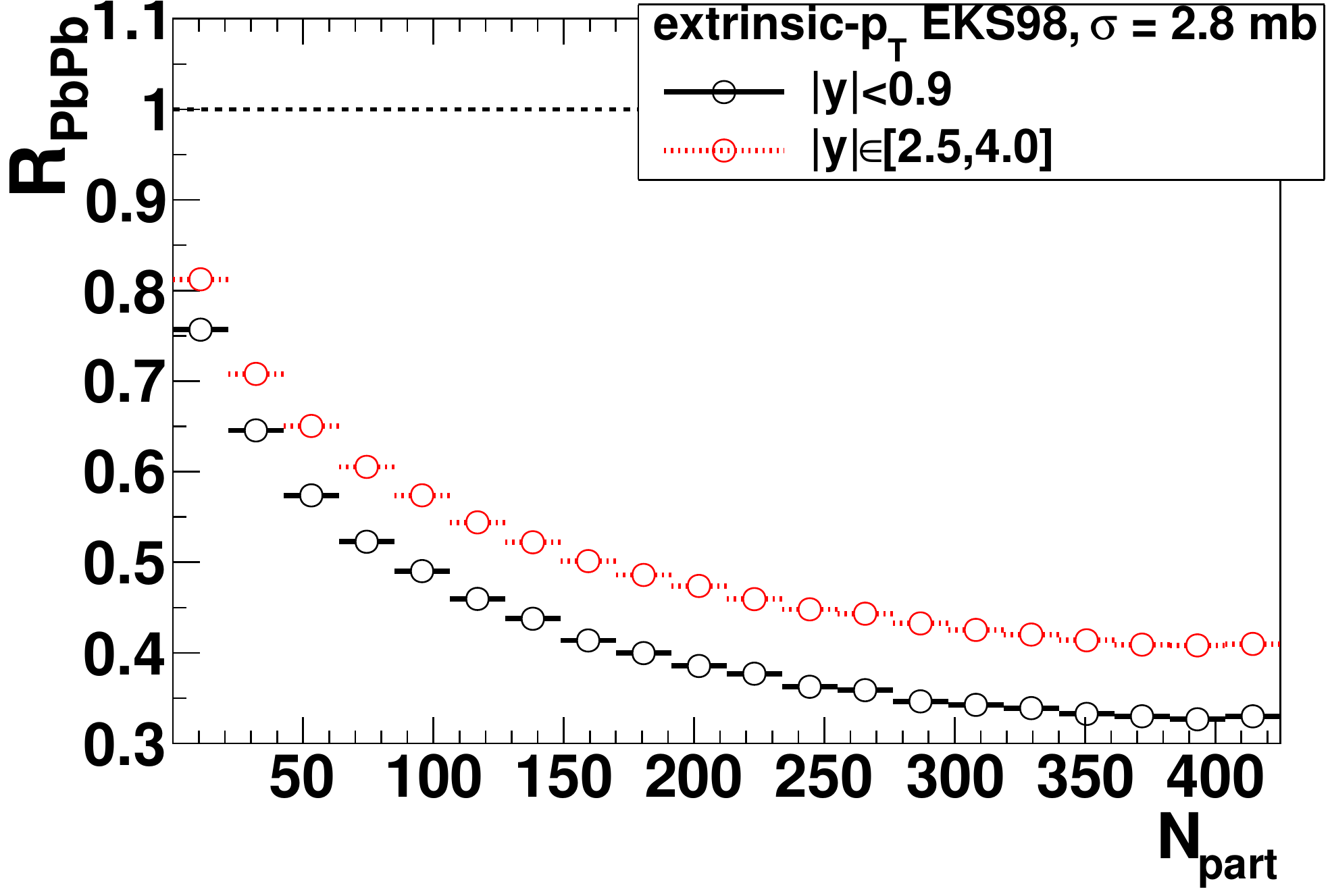}}
\end{center}
\caption{(Color online) \jpsi\ nuclear modification factor, \RPbPb, in PbPb collisions at $\sqrt{s_{NN}}=5.5\mathrm{~TeV}$ versus \Npart, using EKS98~\cite{Eskola:1998df} gluon shadowing parametrisation and three values of the nuclear absorption cross section in the extrinsic scheme. \RPbPb\ is shown for two different experimental acceptances in rapidity, $|y|<0.9$ and $|y|\in [2.5, 4]$.}
\label{fig:RPbPb_vs_Npart}
\end{figure*}

In the following, we present our results for the \jpsi\ nuclear modification factor due to CNM effects in the extrinsic sheme in \pPb\ and PbPb collisions at $\sqrt{s_{NN}} = 5.5\mathrm{~TeV}$: $R_{AB} = dN_{AB}^{J/\psi}/\langle\Ncoll\rangle dN_{pp}^{J/\psi}$, where $dN_{AB}^{J/\psi} (dN_{pp}^{J/\psi})$ is the observed \jpsi\ yield in $AB = \pPbm,\PbPbm$ (\pp) collisions and $\langle\Ncoll\rangle$ is the average number of nucleon-nucleon collisions occurring
in one \pPb\ or PbPb collision. Without nuclear effects, $R_{AB}$ should equal unity.

In \cf{fig:RpPb_vs_y_and_RPbPb_vs_y_vs_cent-a}, we show $R_{\pPbm}$ versus $y$. The curve with no absorption allows us to highlight the strong rapidity dependence of the shadowing. We can also notice that the shadowing alone should already be responsible for a quite large amount of \jpsi\ suppression, up to \mbox{$36\,\%$} at $y=6$. This is expected due to the very small $x$-region in the gluon nPDF that becomes accessible at LHC energy (down to $10^{-5}$). At backward rapidity, we are in the antishadowing region, with the antishadowing peak at $y \simeq -5$. \cf{fig:RpPb_vs_y_and_RPbPb_vs_y_vs_cent-b} shows that the $y$-dependence of $R_{\PbPbm}$ for the centrality bins used by the ATLAS Collaboration \cite{aad} is similar for all the bins, with a dip at mid-$y$. This shape is the opposite of the one obtained at RHIC energy~\cite{Ferreiro:2008wc,Ferreiro:2009ur}, with a peak at mid-$y$. 
One can also notice that $R_{\PbPbm}$ decreases at very large rapidity showing that one has gone beyond the antishadowing peak. Here, $R_{\PbPbm}$ is systematically smaller at mid-$y$ than at forward-$y$. This is also illustrated on \cf{fig:RPbPb_vs_Npart}, with the centrality dependence of $R_{\PbPbm}$ for two regions in~$y$.
This behaviour of the CNM effects may partially -- or completely -- compensate the opposite effect expected from \ccbar\ recombination, with a maximum enhancement at $y=0$. Overall, one may observe a $R_{\PbPbm}$ rather independent of $y$ resulting of two $y$-dependent effects.

N.M. acknowledges financial support from  F.R.S.-FNRS (Belgium). E.G.F. thanks Xunta de Galicia
(2008/012) and Ministerios de Educacion y Ciencia of Spain
(FPA2008-03961-E/IN2P3) for financial support.





\bibliographystyle{elsarticle-num}



\end{document}